\documentclass[english]{revtex4}
\usepackage[T1]{fontenc}
\usepackage[latin9]{inputenc}
\usepackage{babel}
\begin{document}

\title{On the properties of the irrotational dust model}
\author{J\k{e}drzej \'Swie\.zewski}
\email{swiezew@fuw.edu.pl}
\affiliation{Faculty of Physics, University of Warsaw, Ho\.za 69, 00-681 Warszawa, Poland}

\begin{abstract}
In this note we analyze the model of the irrotational dust used recently to deparametrize gravitational action. We prove that the remarkable fact that the Hamiltonian is not a square root is a direct consequence of the time-gauge choice in this model. No additional assumptions or sign choices are necessary to obtain this crucial feature. In this way we clarify a point recently debated in the literature.

\end{abstract}

\maketitle
\section{Introduction}

The possibilty of obtaining a simpler, e.g., avoiding the problem of time, description of gravitational interactions by the means of a coupling to specific matter fields has been present in the literature
for some time \cite{KucharTorre, RovelliSmolin, BrownKuchar, BicakKuchar}. Some of those models gained more attention due to new possibilities they provided for the quantization
of gravitational interaction especially in the context of Loop Quantum Gravity. One of the models considered is the irrotational dust model first introduced in \cite{BrownKuchar}. It has recently been analyzed by the authors of \cite{VHTP, Cosmo, proceed}, where it is claimed that the surprising fact that the Hamiltonian of this theory is not a square root, together with the kinematical
structure of Loop Quantum Gravity, provides a complete theory of quantum gravity.\footnote{See the abstract of \cite{VHTP}.} However, the lack of the square root in this model has been questioned in \cite{KGTT}, where it is claimed that the square root may not explicitly appear in the action due to an artificial sign choice, but removing it in this way leads to serious problems on the quantum level. Hence, it is stated there that the square root is present in the theory in the form of an absolute value (a square root of a square). The aim of this note is to clarify the issue of whether the square
root (or the absolute value) is absent from the model of irrotational dust and whether any artificial sign choices are necessary for this important feature to take place.

\section{The irrotational dust model}

The theory analyzed here consists of the irrotational dust field $T$
coupled to gravity. It is described by the following action
\begin{equation}
S=S_{GR}+S_{SM}+S_{D}=\int d^{4}x\sqrt{-\det g}R+\int d^{4}xL_{SM}+\int d^{4}xL_{D},\label{eq:action}
\end{equation}
where the first term is the Hilbert-Einstein action, the second describes any type of standard matter content and the last term is the dust action. The dust Lagrangian is of the form
\begin{equation}
L_{D}=-\frac{1}{2}\sqrt{-\det g}M(g^{\mu\nu}\partial_{\mu}T\partial_{\nu}T+1),\label{eq:dustlagrangian}
\end{equation}
where the non-dynamical field $M$ plays a role of a Lagrange multiplier.
In the following we will discuss the deparametrization scheme of the
action (\ref{eq:action}) with respect to the dust field $T$, in
which the rest of the matter content plays no role, hence we disregard it in
the remainder of this work.

After introducing ADM \cite{ADM} variables to describe the
geometric degrees of freedom, namely
\begin{equation}
g_{\mu\nu}=\left(\begin{array}{cc}
-N^{2}+N^{a}N_{a} & N_{a}\\
N_{a} & h_{ab}
\end{array}\right),
\end{equation}
one finds the dust Lagrangian (\ref{eq:dustlagrangian}) in the form
\begin{equation}
L_{D}=\frac{1}{2}\sqrt{\det h}\frac{M}{N}\big((\dot{T}-N^{a}\partial_{a}T)^{2}-N^{2}(h^{ab}\partial_{a}T\partial_{b}T+1)\big).
\end{equation}
Introducing the momentum conjugate to the dust field
\begin{equation}
p_{T}=\frac{\partial L_{D}}{\partial\dot{T}}=\frac{\sqrt{\det h}M}{N}(\dot{T}-N^{a}\partial_{a}T)\label{eq:ped}
\end{equation}
and rewriting the dust Lagrangian with the use of the momentum, the
Legendre transformation can be completed. The dust action has the
form
\begin{eqnarray}
S_{D}=\int dtd^{3}x\bigg(\dot{T}p_{T}-N\left(\frac{1}{2}\left(M\sqrt{\det h}\right)^{-1}p_{T}^{2}+\frac{1}{2}M\sqrt{\det h}\left(h^{ab}\partial_{a}T\partial_{b}T+1\right)\right)-N^{a}\left(p_{T}\partial_{a}T\right)\bigg).\label{eq:actiondustpart}
\end{eqnarray}
Now one can proceed along two paths.

\subsection{Previous treatment}

The authors of \cite{VHTP, Cosmo, proceed} and the
authors of \cite{KGTT} turn to the analysis of the equations
of motion implied by that action. The equation for $M$ reads
\begin{equation}
M^{2}=\frac{p_{T}^{2}}{\det h\ (1+h^{ab}\partial_{a}T\partial_{b}T)}.\label{eq:Msquare}
\end{equation}
This condition links the values of $M$ and $p_{T}$, but not their
signs. In order to express $M$ as a function of $p_{T}$ the authors
of \cite{VHTP, Cosmo, proceed} invoke an argument
about the role of energy density played by $M$ in the stress-energy
tensor of the dust. Using this additional requirement they guarantee
positivity of $M$.\footnote{This argument is most clearly stated in \cite{proceed}.} Hence they can solve (\ref{eq:Msquare}) and obtain
\begin{equation}
M=\frac{\left|p_{T}\right|}{\sqrt{\det h}\sqrt{1+h^{ab}\partial_{a}T\partial_{b}T}}.
\end{equation}
This expression is then plugged into the dust action and the following
expression is obtained
\begin{equation}
S_{D1}=\int dtd^{3}x\left(\dot{T}p_{T}-N\left|p_{T}\right|\sqrt{1+h^{ab}\partial_{a}T\partial_{b}T}-N^{a}\left(p_{T}\partial_{a}T\right)\right).
\end{equation} 
Note that an absolute value of $p_{T}$ is present in the action.
To dispose of the absolute value the authors of \cite{VHTP,
Cosmo, proceed} fix the sign of $p_{T}$ to be positive by hand
and proceed with their treatment. This sign choice is of crucial importance
to the present note and we will come back to it later. At this stage
the authors introduce the time-gauge namely they choose coordinates
such that
\begin{equation}
t=T.
\end{equation}
Two simple consequences of this choice are $\dot{T}=1$ and $\partial_{a}T=0$.
Moreover, the dynamical preservation of this gauge enforces the condition
\begin{equation}
N=1.
\end{equation}
Implementing these facts, they obtain the full constraints of the theory in the form
\begin{eqnarray}
C^{\rm tot}=C+p_T=0 \label{constraint},\\
C^{\rm tot}_a=C_a=0,
\end{eqnarray}
where $C$ and $C_a$ are the standard Hamiltonian and vector constraints of canonical gravity. Solving the first constraint for $p_T$, they complete the deparametrization of the
action with respect to the dust ending up with a theory given by the
action
\begin{equation}
S_{dep}=\int dtd^{3}x\Big(\dot{h}_{ab}\pi^{ab}-C-N^{a}C_{a}\Big),\label{eq:finalaction}
\end{equation}
where $C$ plays a role of a true, nonvanishing, Hamiltonian of the theory generating
evolution in the dust time.

This result has been questioned in \cite{KGTT}, where
the authors argue that the sign choice of $p_{T}$ imposes a sign
choice of $C$ (due to the constraint (\ref{constraint})), hence although no absolute value appears explicitly in (\ref{eq:finalaction}), it is there implicitly since the sign
of $C$ is limited. This is then argued to be a source of problems in the
process of quantization of the considered theory, since imposing the
sign condition on the quantum level requires a detailed knowledge
of the spectrum of $C$, which is not currently available.

\subsection{Alternative treatment}

Arriving at (\ref{eq:actiondustpart}), one can follow a different
route. Instead of analyzing stationarity of the action with respect
to variations of $M$ to express it as a function of $p_{T}$, one
can realize one more consequence of the choice of the time-gauge.
As already noticed in \cite{VHTP} the choice $t=T$
leads to $\dot{T}=1$, $\partial_{a}T=0$ and $N=1$. These results
implemented into the definition of the dust momentum, given by (\ref{eq:ped}), imply
\begin{equation}
p_{T}=\sqrt{\det h}M. \label{eq:pedzacy}
\end{equation}
Note that here we can express $M$ as a function of $p_{T}$ without
any choices of signs. The $M$ obtained from this equality satisfies
the stationarity condition (\ref{eq:Msquare}), and when plugged into
(\ref{eq:actiondustpart}) leads to
\begin{equation}
S_{D2}=\int dtd^{3}x\left(\dot{T}p_{T}-Np_{T}\sqrt{1+h^{ab}\partial_{a}T\partial_{b}T}-N^{a}\left(p_{T}\partial_{a}T\right)\right),
\end{equation}
where again the spatial derivatives of $T$ vanish due to the gauge
choice. After solving the constraints as it was done previously, we end up with a theory given by the action functional apparently identical with (\ref{eq:finalaction}), with the crucial difference
that the sign of $C$, being linked with the sign of $p_{T}$ by (\ref{constraint}), is
no longer limited since the latter sign is not fixed in the presented treatment.

If one wishes to impose the condition limiting $M$ to be positive introduced in \cite{VHTP, Cosmo, proceed} then from (\ref{eq:pedzacy}) we see that it limits $p_T$ to be positive and from (\ref{constraint}) also $C$ to be negative. We now see clearly that it is actually the stress-energy condition imposed on $M$ which limits the sign of $C$ and that the strongest result is obtained if no sign choices are introduced.

\section{Summary}

To summarize, we analyzed the question whether gravitional action
deparametrized with the use of the irrotational dust possesses the feature
of its Hamiltonian not being a square root. The importance of this
feature has been underlined in both the works of \cite{VHTP,
Cosmo, proceed} and \cite{KGTT}, however, the latter
work criticizes the price that is paid to obtain it in the former
treatment. Here we showed that this crucial feature is a direct consequence
of the time-gauge in this model. Hence, no additional input, like
the stress-energy tensor argument or the artificial sign choice is
necessary. Therefore, at least some of the problems of the model pointed out
in \cite{KGTT} can be avoided. We leave the evaluation of
the full consequences of the new treatment on the quantum level for
future research.

\end{document}